\documentstyle[multicol,aps,prl,epsfig]{revtex}

\begin{document}
\title{A modest proposal to solve the ''missing mass'' problem and related
cosmological paradoxes}
\author{Julio A. Gonzalo}
\address{Facultad de Ciencias\\
Universidad At\'onoma de Madrid\\
Cantoblanco 28049 Madrid Spain\\
email: julio.gonzalo@uam.es}
\date{24-10-2000}
\maketitle

\begin{abstract}
Properly interpreted data from nearby galaxies $(z\simeq 0.01)$ lead to $%
\Omega \simeq 0.082.$ Data from farther away galaxies $(z\simeq 1)$ with
type Ia supernovae to $\Omega =0.153.$ Data to be expected from very high
redshifted galaxies $(z\simeq 10.1)$ to $\Omega =0.500.$ And actual data
from the CBR, emitted at the time at which the universe became transparent $%
(z\simeq 1422)$ to $\Omega \simeq 0.992$. All these data are simultaneously
consistent with the standard big-bang picture (no inflation), in which $%
\Omega $ is time dependent and it is given by $\Omega (y)=1/\cosh ^{2}(y)$,
being $y\equiv \sinh ^{-1}(T_{+}/T)^{1/2}$
\end{abstract}



\vskip1pc

\begin{multicols}{2}
\narrowtext
\parskip=0cm

It is well known \cite{weinberg,peebles} that Einstein%
\'{}%
s cosmological equations for {\bf an open universe }can be written as 
\begin{equation}
\stackrel{\circ }{R}=R^{-1/2}\left\{ (8\pi /3)G\rho R^{3}+\left| k\right|
c^{2}R\right\} ^{1/2}  \label{R}
\end{equation}

where $R(t)$ is the scale factor or radius of the universe, $\stackrel{\circ 
}{R}(t)$ its time derivative, $\rho (t)$ the mass density per unit volume,
being $\rho (t)=\rho _{m}(t)+\rho _{r}(t)$ the sum of matter $(\rho _{m})$
and radiation $(\rho _{r})$ mass densities, and $k<0$ the spatial curvature.

In the present matter dominated epoch $\rho _{m}>>\rho _{r}$ and $\rho \sim
R^{-3}$. Then we take $\rho R^{3}\sim const$ in the right hand side of Eq.%
\ref{R} and define 
\begin{equation}
(8\pi /3)G\rho _{+}R_{+}^{3}=\left| k\right| c^{2}R_{+}  \label{8pi}
\end{equation}

to write the integral as 
\begin{equation}
\int dt=\int \frac{R^{1/2}}{\left\{ (8\pi /3)\rho _{+}R_{+}^{3}+\left|
k\right| c^{2}R\right\} ^{1/2}}dR  \label{intg}
\end{equation}

where $(8\pi /3)G\rho _{+}R_{+}^{3}=\left| k\right| c^{2}R_{+}=a^{2}$ is a
constant. Using the change of variable $x^{2}=\left| k\right| c^{2}R$ the
integral in the right hand side of Eq.\ref{intg} is transformed to 
\[
\int \frac{x^{2}}{\left\{ a^{2}+x^{2}\right\} }dx, 
\]

which is found in tables, resulting in 
\begin{equation}
t=\frac{R_{+}}{\left| k\right| ^{1/2}c}\left\{ \sinh (y)\cosh (y)-y\right\} ,
\label{t}
\end{equation}

where $(R/R_{+})^{1/2}\equiv \sinh (y)$ has been used, i.e. 
\begin{equation}
R=R_{+}\sinh ^{2}(y)  \label{Rsin}
\end{equation}

Eqs.\ref{t} and \ref{Rsin} give parametrically, in terms of $y$, the {\bf %
cosmic time }$t(y)$ and the {\bf cosmic radius }$R(y)$. Using Eqs.\ref{t}
and \ref{Rsin} expressions for the Hubble parameter $H\equiv \stackrel{\circ 
}{R}/R$ and the dimensionless cosmic parameters $(Ht)$ and $\Omega =(\rho
/\rho _{c})$ (where $\rho _{c}=3H^{2}/8\pi G$ is the critical density), all
of them time dependent, can be easily obtained. In particular 
\begin{equation}
H=(c\left| k\right| ^{1/2}/R_{+})\cosh (y)/\sinh ^{3}(y)  \label{H}
\end{equation}

resulting in 
\begin{equation}
(Ht)=\left\{ \sinh (y)\cosh (y)-y\right\} \cosh (y)/\sinh ^{3}(y)\geq 2/3
\label{Ht}
\end{equation}
\begin{equation}
\Omega =1/\cosh ^{2}(y)\leq 1  \label{omega}
\end{equation}

Improved recent determinations \cite{freedman} of $H_{0}\simeq 65\pm
10kmsec^{-1}Mpc^{-1}$ (mainly from nearby galaxies) and \cite{botte} $%
t_{0}\simeq (13.7\pm 2)\times 10^{9}$ years (from the oldest stars in the
Milky Way globular clusters) result in 
\begin{equation}
H_{0}t_{0}\simeq 0.91\text{ (dimensionless)}  \label{Ht2}
\end{equation}

which , by Eq.\ref{Ht}, implies $y_{0}\simeq 1.92$. Then, taking into
account that the corresponding cosmic {\bf equation of state} $RT\simeq $%
constant, 
\begin{equation}
y_{0}=\sinh ^{-1}(R_{0}/R_{+})^{1/2}=\sinh ^{-1}(T_{+}/T_{0})^{1/2}\simeq
1.92,  \label{ycero}
\end{equation}

which allows one to get $T_{+}\simeq 30.4K$, using the well known COBE data 
\cite{mather} for $T_{0}\simeq 2.726\pm 0.01K$, and, consequently, to
evaluate cosmic quantities at any $R=R_{0}/(1+z)$ and $T=T_{0}(1+z)$, being $%
z$ the redshift.

First let us determine $z_{+}$ corresponding to $T_{+}=30.4K$ (i.e. to $%
R_{+} $ defined by Eq.\ref{8pi}). By means of Eq.\ref{ycero} we get 
\begin{equation}
y_{0}\simeq \sinh ^{-1}\left( \frac{1+z_{+}}{1+0}\right)
=1.92\longrightarrow z_{+}=10.1  \label{ycero2}
\end{equation}

This implies $y_{+}=\sinh ^{-1}(1)=0.881$ for $z_{+}=10.1$, which
corresponds to a density parameter $\Omega _{+}=1/\cosh ^{2}(y_{+})=1/2.$

We can evaluate the evolution of the expected value of the density parameter 
$\Omega (z)$ for $(1)$ $z\simeq 0.01$, {\bf nearby galaxies}, $(2)$ $z\simeq
1$, relatively {\bf distant galaxies} (accurately characterized \cite{astro1}
recently by means of Type Ia Supernovae), $(3)$ $z\simeq 10.1$, a guess on
the upper redshift for far away {\bf protogalaxies}, and $(4)$ $z\simeq 1422$%
, corresponding to the distance at which {\bf cosmic background fluctuations}
are originated, i.e. the distance at which the universe is becoming
transparent and therefore $T_{af}\simeq 3880K$ (i.e. $z=(T_{af}/T_{0})-1%
\simeq 1422$).

Then we have

$(1)\qquad z=0.01$ ({\bf nearby galaxies})

$\qquad \qquad y_{z}=\sinh ^{-1}\left( \frac{1+z_{+}}{1+z}\right) \simeq
1.913\longrightarrow \Omega =1/\cosh ^{2}(y_{z})\simeq 0.083$

$(2)\qquad z=1$ (relatively {\bf distant galaxies})

$\qquad \qquad y_{z}=\sinh ^{-1}\left( \frac{1+z_{+}}{1+z}\right) \simeq
1.592\longrightarrow \Omega =1/\cosh ^{2}(y_{z})\simeq 0.152$

$(3)\qquad z=10.1$ ({\bf guess} for {\bf protogalaxies})

$\qquad \qquad y_{z}=\sinh ^{-1}\left( \frac{1+z_{+}}{1+z}\right) \simeq
0.881\longrightarrow \Omega =1/\cosh ^{2}(y_{z})\simeq 0.500$

$(4)\qquad z=1422$ ({\bf CBR}, {\bf the very first light})

$\qquad \qquad y_{z}=\sinh ^{-1}\left( \frac{1+z_{+}}{1+z}\right) \simeq
0.0882\longrightarrow \Omega =1/\cosh ^{2}(y_{z})\simeq 0.992$

This means that the standard big-bang model (no inflation) can account {\bf %
directly} and {\bf simultaneously} for the various observed $\Omega $
values, from $\Omega \simeq 0.1$ for light coming from our immediate
neighborhood, all the way to $\Omega \simeq 1$ for light coming from the
moment at which the universe became transparent. Here, no {\bf ''missing
mass''}. No need for {\bf ''inflation''} to explain the ''missing mass''. No
need for a {\bf non-vanishing cosmological constant} to account for the
apparent redshift dependence of distant \ cosmic objects.

The recent reported findings of the Boomerang Collaboration \cite{astro2} $%
(\Omega =1.06\pm 0.06)$ and the Maxima Collaboration $(\Omega =0.90\pm 0.07)$%
, obtained by means of the analysis of the power spectrum of spatial
temperature fluctuations from spherical harmonic fits to their CMB maps, are
in very good agreement with our estimate \cite{cereceda} $(\Omega =0.992)$
of the cosmic density parameter {\bf at the time the CBR was emitted}.

\end{multicols}
\end{document}